\documentclass[reprint,prx,floatfix,twocolumn,superscriptaddress]{revtex4-2}

\usepackage{amsmath}
\usepackage{amssymb}
\usepackage{float}
\usepackage{graphicx}
\usepackage{hyperref}
\usepackage{lipsum}
\usepackage{siunitx}
\usepackage[version=4]{mhchem}

\graphicspath{ {./figures/} {./figures_supplement/} }

\newcommand{\Eref}[1]{Eq.~\ref{#1}}

\newcommand{\Fref}[1]{Fig.~\ref{#1}}

\makeatletter

\usepackage{dcolumn}
\usepackage{bm}
\usepackage{float}
\makeatother

\begin{document}

\title{
Superconducting Qubits Above 20 GHz Operating over 200 mK
}

\author{Alexander Anferov}
\email{aanferov@uchicago.edu}
\affiliation{James Franck Institute, University of Chicago, Chicago, Illinois 60637, USA}
\affiliation{Department of Physics, University of Chicago, Chicago, Illinois 60637, USA}

\author{Shannon P. Harvey}
\affiliation{Department of Applied Physics, Stanford University, Stanford, California 94305, USA}
\affiliation{SLAC National Accelerator Laboratory, Menlo Park, CA, 94025 USA}

\author{Fanghui Wan}
\affiliation{Department of Applied Physics, Stanford University, Stanford, California 94305, USA}
\affiliation{SLAC National Accelerator Laboratory, Menlo Park, CA, 94025 USA}

\author{Jonathan Simon}
\affiliation{Department of Applied Physics, Stanford University, Stanford, California 94305, USA}
\affiliation{Department of Physics, Stanford University, Stanford, California 94305, USA}

\author{David I. Schuster}
\email{dschus@stanford.edu}
\affiliation{Department of Applied Physics, Stanford University, Stanford, California 94305, USA}
\affiliation{SLAC National Accelerator Laboratory, Menlo Park, CA, 94025 USA}

\date{\today }

\begin{abstract}
Current state-of-the-art superconducting microwave qubits are cooled to extremely low temperatures to avoid sources of decoherence.
Higher qubit operating temperatures would significantly increase the cooling power available, which is desirable for scaling up the number of qubits in quantum computing architectures and integrating qubits in experiments requiring increased heat dissipation.
To operate superconducting qubits at higher temperatures, it is necessary to address both quasiparticle decoherence (which becomes significant  for aluminum junctions above 160~mK) and dephasing from thermal microwave photons (which are problematic above 50~mk).
Using low-loss niobium trilayer junctions, which have reduced sensitivity to quasiparticles due to niobium's higher superconducting transition temperature, we fabricate transmons with higher frequencies than previously studied, up to 24~GHz.
We measure decoherence and dephasing times of about $1~\mu$s, corresponding to average qubit quality factors of approximately $10^5$, and find that decoherence is unaffected by quasiparticles up to $1~$K.
Without relaxation from quasiparticles, we are able to explore dephasing from purely thermal sources, finding that our qubits can operate up to approximately $250~$mK while maintaining similar performance.
The thermal resilience of these qubits creates new options for scaling up quantum processors, enables hybrid quantum experiments with high heat dissipation budgets, and introduces a material platform for even higher-frequency qubits.

\end{abstract}

\maketitle
\section{Introduction}
Superconducting qubits built from Josephson junctions are a promising technology for quantum sensing, hybrid systems coupling different types of quantum emitters \cite{clerk2020hybrid,xiang2013hybridrev}, and realizing large-scale quantum computing architectures \cite{kjaergaard2020qubitReview}.
While current qubits typically operate below 10~GHz due to widely available cryogenic microwave equipment, increased frequencies expand the range of accessible energies in quantum experiments, can couple to a new range of signals and quantum emitters, and importantly enable higher operating temperatures.
This offers a straightforward approach for scalability by making use of significantly higher cooling power available with even a moderate increase in temperature \cite{Pobell2007}. 
Thermally-resilient qubits could reduce hardware overhead for microwave quantum interconnects \cite{pechal2017millimeter, magnard2020link}, provide new opportunities for direct integration with superconducting digital logic \cite{liu2023sfqControl,leonard2019sfqControl,mcdermott2014sfqControl}, and help manage the heat load from expanding numbers of qubit control lines
 \cite{Krinner2019qubitLines} as superconducting quantum processors \cite{Wu2021qproc,Arute2019qproc,Kim2023qproc} expand beyond hundreds of qubits.
For hybrid quantum systems \cite{clerk2020hybrid,xiang2013hybridrev}, many of which introduce additional challenges by exposing qubits to magnetic fields or direct optical illumination \cite{kumar2023quantum, wang2022high, mirhosseini2020superconducting, jiang2020efficient}, more resilient qubits could also help reduce experiment complexity and improve performance. 

Increasing qubit operating temperature requires a two-pronged approach addressing both increased environment radiation as well as quasiparticle loss.
Specifically, thermal photons at the qubit frequency result in heating, requiring active reset \cite{han2023active,Magnard2018reset} or extensive dissipation engineering \cite{Wang2021cooling,valenzuela2006cooling} to artificially cool the qubit to its ground state before experiments begin.
Furthermore, thermal population in the readout resonator (less easily cooled \cite{Wang2019cavitycooling}) induces qubit dephasing \cite{Clerk2007dephasing}.
Increasing system frequencies mitigates these effects due to the reduced thermal sensitivity of higher energy photons, and expands accessible operating temperatures and cooling powers even for conventional aluminum Josephson junction qubits \cite{mcdermott2009materialDecoherence,kjaergaard2020qubitReview}.

Ultimately operating temperatures are limited by decoherence from quasiparticles \cite{connolly2023nonequilib,martinis2009nonequilib,Serniak2018hotquasiparticle,catelani2011quasiparticle,catelani2014quasiparticle,mattis1958bardeen} in the superconductor.
Niobium's higher energy gap relative to aluminum results in significantly lower quasiparticle density at elevated temperatures as well as increased nonequilibrium quasiparticle recombination rates \cite{leo2011nbquasiparticle}, making it a promising material for realizing higher-temperature qubits and increasing repetition rates for optical transduction experiments \cite{kumar2023quantum, wang2022high, mirhosseini2020superconducting, jiang2020efficient}.
Full temperature resilience requires replacing aluminum junctions with a high-temperature nonlinear element, which can be realized with newly optimized niobium nitride \cite{kim2021nbnjj} or niobium-based \cite{anferov2023nbjj} junctions: along with increased plasma frequencies \cite{Petley1969josephson} that simplify high-frequency qubit design, these junctions have shown rapidly improving coherence properties in microwave qubits and reduced quasiparticle sensitivity \cite{anferov2023nbjj}.

In this work, we scale up transmon qubits to 11--24 GHz, expanding the frequency range studied by superconducting devices.
Using micron-scale optically-defined niobium trilayer Josephson junctions, we achieve qubit dephasing times above $1~\mu$s through material improvements and residue removal techniques.
With higher frequencies and niobium's quasiparticle insensitivity we improve qubit thermal resilience, demonstrating coherent behavior above 200~mK, and investigate their coherence properties up to 1~K in a previously unexplored regime.

\section{High Frequency Qubit Design}
\begin{figure}
\centering
\includegraphics[width=3.37in]{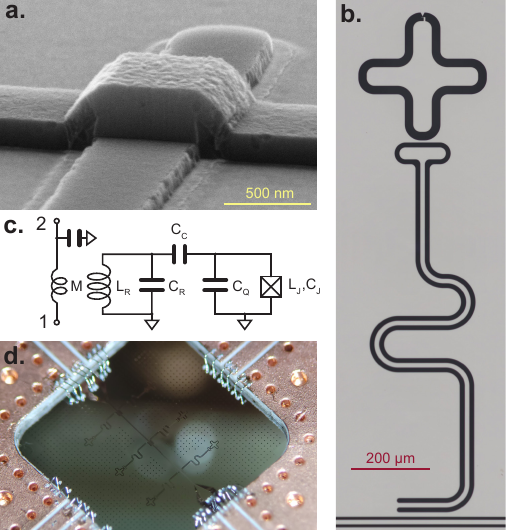}
\caption{
Qubit geometry.
(a) A scanning electron micrograph of a low-loss niobium trilayer junction at the core of the qubit.
(b) A micrograph of the qubit and readout resonator geometry, with the junction location marked at the top.
(c) Equivalent circuit of the qubit and readout resonator coupled inductively to a transmission line.
(d) Photograph of a chip containing six qubits mounted in a low-loss K band circuit board.
\label{fig1}}
\end{figure}
The key element at the heart of our transmon qubit is a niobium trilayer Josephson junction with high critical current density ($J_c$), shown in \Fref{fig1}a.
We use a self-aligned fabrication process (see Appendix \ref{appendix:fab}) to form a Nb/Al/AlO$_x$/Al/Nb Josephson junction on a C-plane sapphire substrate, with two main changes from Ref. \cite{anferov2023nbjj}. 
First, we use low-temperature plasma-enhanced chemical vapor deposition (PECVD) for growing the \ce{SiO2} spacer \cite{gronberg2017swaps}: keeping the wafer below $230~^\circ$C during the growth process creates a spacer oxide with lower loss than other methods such as high-density PECVD\cite{anferov2023nbjj} (see Appendix B for more detail), and preserves a high $J_c$ density needed for higher-frequency junctions.
Second, immediately prior to measurement we use a \qty{72\pm1}{\degreeCelsius} solution of catechol, hydroxylamine and 2-(2-aminoethylamine)-ethanol (Dupont EKC 265) to remove lossy plasma etch residues along with a thin layer of the oxidized niobium surface, known to contain lossy oxides \cite{verjauw2021nbOxide,premkumar2021nbOxide}.
This treatment (see Appendix \ref{appendix:ekc}) leaves a smooth metal surface with minimal defects and residues. 
For these PECVD junctions, the $J_c$ density can be adjusted between 0.1--$2.6~\text{kA}/\text{cm}^{2}$ (see Appendix \ref{appendix:pecvd}).
Although higher densities are possible, we select a $J_c$ between 0.19--$0.36~\text{kA}/\text{cm}^{2}$ which is closer to typical aluminum junctions (typically $0.05~\text{kA}/\text{cm}^{2}$ \cite{serniak2019nonequilibrium}), allowing us to design qubits between 11--24~GHz using junction finger dimensions between 0.4--$0.8~\mu$m, achieved entirely with optical lithography.
We note that junction reproducibility and minimum feature size could also be improved with electron beam lithography. 

Our qubit geometry, shown in \Fref{fig1}b, is qualitatively similar to conventional microwave transmon qubits \cite{koch2007cpb,barends2013xmon}, with the primary difference that every major feature is slightly smaller.
The rounded cross-shaped qubit capacitor $C_Q$ is capacitively coupled to a 20--22~GHz quarter-wave coplanar waveguide resonator used for dispersive measurements.
The other end of the resonator is inductively coupled ($\kappa/2\pi= 2$--9~MHz) to a common feedline.
The system can be modelled by the circuit in \Fref{fig1}c, where the qubit frequency is determined by the Josephson inductance $L_J$ and total shunt capacitance $C_\Sigma$.
As with linear circuits, reducing capacitance and inductance increases qubit frequency.
Smaller capacitance has the benefit of increased anharmonicity and thus faster possible qubit control operations, which can help improve single-qubit gate fidelities.
While this speedup could be exploited until the device begins to approach the charge-sensitive regime \cite{Serniak2018hotquasiparticle}, because the qubits reported in this manuscript use optical lithography, they have a large capacitance that limits the anharmonicity and therefore gate speeds. However, designing 20~GHz qubits with lower capacitance will enable substantially faster gate times than achievable with microwave qubits.

The qubit capacitor, ground plane and readout resonator are defined along with the junction, so no additional fabrication steps are needed.
Chips with several qubits and their readout resonators sharing a common microwave feedline are patterned on 5~mm square chips, and mounted in a K-band (18-27~GHz) package shown in \Fref{fig1}d, which is carefully engineered for low-loss and mode-free operation up to 30~GHz (see Appendix \ref{appendix:packaging}).
The assembly is thermalized to the base stage of a dilution refrigerator (65-95~mK), where transmission measurements through the central feedline can be used to control and measure the qubits \cite{schuster2005ac, koch2007cpb}.

\section{Qubit Characterization}
\begin{figure*}
\centering
\includegraphics[width=6.67in]{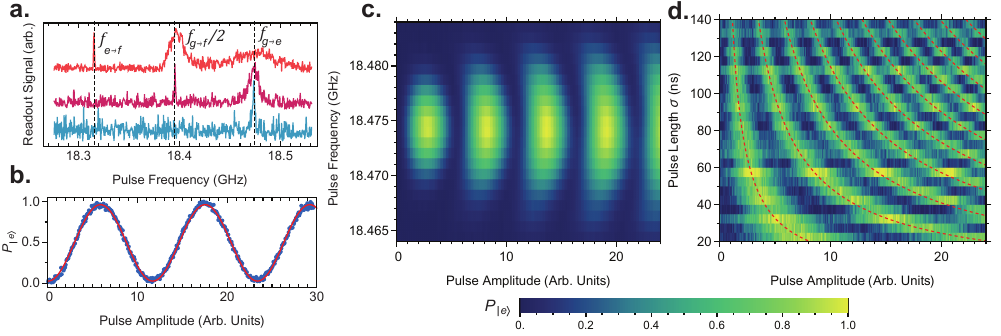}
\caption{
Qubit dynamics.
(a) Deflection of the readout resonator transmission signal as a function of applied qubit pulse frequency, shown for increasing qubit pulse power.
At low powers (blue) a single peak is observed when the pulse is resonant with the qubit frequency ($f_{ge}=18.474~$GHz).
As power increases, the linewidth of this transition increases, and additional peaks appear from excitations into higher qubit levels through many-photon excitations ($f_{gf}/2$ etc).
These features have a spacing of $\alpha/2 = (f_{ge}-f_{ef})/2$.
(b) Measured excited state probability shows Rabi oscillations when a fixed-length pulse with varying amplitude is applied at the qubit frequency.
The red line is a fit to the expected sinusoidal behavior.
(c) Rabi oscillations are measured for frequencies near $f_{ge}$, with brighter colors corresponding to higher excited state probabilities.
Away from the transition frequency, the Rabi frequency increases while the oscillation amplitude decreases and becomes power dependent.
(d) Rabi oscillations as a function of pulse amplitude and length $\sigma$, with brighter colors corresponding to higher excited state probabilities.
Dashed red lines mark contours of integer $\pi$ pulses where $\sigma\Omega=m\pi$.
\label{fig:fig2}}
\end{figure*}

The qubit properties are experimentally determined with microwave spectroscopy \cite{paik2011qubit3d,koch2007cpb}.
While monitoring low-power transmission through the system at the readout resonator frequency, simultaneously applying a second probe pulse reveals the energy spectrum of the qubit, as shown in \Fref{fig:fig2}a.
At low powers, we observe a deflection in transmission at the bare qubit frequency $f_{ge}$ as the qubit is excited, resulting in a dispersive shift of the resonator frequency. 
Increasing the power of the probe pulse reveals the higher energy states of the qubit through two-photon processes $f_{gf}=(f_{ge}+f_{ef})/2$ and excited state transitions $f_{ef}$.
This allows us to measure the anharmonicity $\alpha \equiv f_{ef}-f_{ge}$, which for our qubits is typically near $200~$MHz.
Notably, this level spacing (which sets an upper bound on qubit gate speed) is similar to many conventional microwave qubits \cite{paik2011qubit3d,koch2007cpb,barends2013xmon}, but could easily be adjusted in our design by picking a different junction $J_c$ density or capacitor size.

\subsection{Time Domain Measurements}
Next we control the qubit. 
We apply fixed-length ($\sigma=25~$ns) Gaussian pulses at the qubit frequency $f_{ge}$ with varying amplitude (over which we have much finer control than time) which results in sinusoidal behavior as shown in \Fref{fig:fig2}b.
We repeat this measurement with $\sigma=60~$ns while varying pulse detuning from the transition $\Delta=f-f_\text{ge}$, and summarize the results in \Fref{fig:fig2}c, where brighter colors indicate the system in the excited state.
At the qubit frequency, we observe a series of bright fringes; however as the detuning from the transition increases, the oscillation rate increases, while the oscillation amplitude is quickly suppressed.
The bandwidth of the fringes is further reduced by the finite nature of the pulse envelope \cite{berman1998finite,fischer2017finite,Boradjiev2013finite}.

To verify the time-dependence of the Rabi oscillations, we also repeat this measurement at $\Delta=0$ while varying $\sigma$ and plot the results in \Fref{fig:fig2}d.
The observed fringes are evenly spaced for fixed amplitude (vertical slice) or fixed length (horizontal): by fitting the fringe contours we find Rabi rates as high as 100~MHz, and calibrate a qubit $\pi$ control pulse with $\sigma$ between 20--40 ns.
Since the bandwidth of these pulses is still much smaller than the level spacing, our single-qubit gate speeds could be optimized to be significantly shorter \cite{rebentrost2009optimal}.
Further reducing our junction area and qubit capacitance to increase qubit anharmonicity could allow for even faster gate operation.

\subsection{Coherence Properties}
\begin{figure*}
\centering
\includegraphics[width=6.67in]{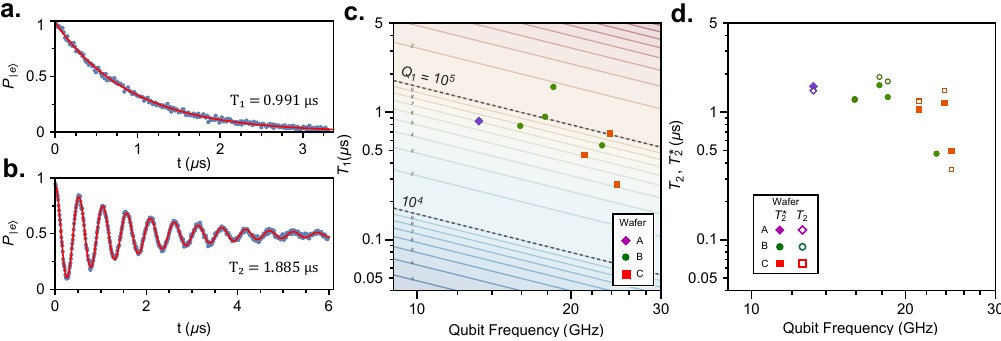}
\caption{
Qubit coherence at 60~mK. 
(a) Average qubit decay time $T_1$ extracted by fitting the decay of excited state population.
(b) Ramsey dephasing experiment from which the dephasing time $T_2^*$ is extracted by fitting the exponential decay of oscillations.
(c) $T_1$ plotted as a function of qubit frequency, grouped by wafer. Background color and numbered lines indicate qubit quality factor $Q_1=\omega_q T_1$. We find an overall mean $Q_1$ of $0.792\times10^5$ with some wafer to wafer variation. 
(d) $T_2^*$ (filled points) and Hahn-echo dephasing time $T_2$ (hollow points) for the qubits studied. We find an average $T_2^*$ and $T_2$ of $1.124~\mu$s and $1.357~\mu$s respectively.
\label{fig:fig3}}
\end{figure*}
We can now study qubit relaxation time and dephasing time, which dictate qubit limitations and act as sensitive probes of loss channels.
Fitting the characteristic population relaxation time $T_1$ (as shown in \Fref{fig:fig3}a) for each qubit, we find that $T_1$ roughly scales inversely with frequency as shown in \Fref{fig:fig3}c, with $T_1=\qty{1.6}{\us}$ for our best device.
We find that $T_1$ typically fluctuates with a relative devation of about 6\% from the average value (See Appendix \ref{appendix:time}), likely from coupled two-level-systems \cite{Carroll2022dynamics,Schlor2019fluctuators,Burnett2019benchmark}.
We note that hardware limitations prevent time-domain measurements of some of the qubits with frequencies outside the mixer bands (Wafers A and C) or those overlapping with the readout resonators at $\sim21$~GHz (Wafer B).
The highest-frequency qubits (e.g. Wafer C) are above their readout resonator frequencies ($\sim22~$GHz) so are likely further affected by Purcell loss.

To probe loss channels in detail we use the frequency-independent qubit quality factor $Q_1=2\pi f_q T_1$, which we find for our devices is just below $10^5$.
Our K-band qubit quality factors are comparable to transmons made with a similar junction process, which measure an average $Q_1=2.5\times10^5$ \cite{anferov2023nbjj}.
The slight decrease in our average $Q_1$ could be attributed to slightly higher junction \cite{anferov2023nbjj} and surface participation \cite{wang2015psurf} arising from our smaller qubit capacitor size relative to the junction area.
By comparison, state-of-the-art aluminum-junction microwave qubits now achieve quality factors of several million \cite{mcdermott2009materialDecoherence}. 
Having addressed potential loss from fluorocarbon residues, we expect that our qubit coherence is now likely primarily limited by loss in residual spacer material (see \Fref{fig1}a), and potentially two-level-systems in the relatively large junction barrier region \cite{Zagoskin2006jjtls,Lisenfeld2015jjtls}. 
This suggests that future devices could be directly improved with smaller junctions and larger capacitor geometry, or with improved junction fabrication techniques \cite{kim2021nbnjj} and spacer removal methods \cite{dunsworth2018airbridge}.
Quantum decoherence and material properties in this frequency range are still relatively unexplored, warranting further investigation \cite{Bilmes2020tlsmap} into the nature and sources of loss.

We also perform a Ramsey experiment to measure the dephasing time $T_2^*$, and a Hahn-echo experiment to characterize the spin-echo dephasing time $T_2$ (\Fref{fig:fig3}b).
We find that $T_2^*$ is consistently higher than $T_1$, nearing the dissipation limit $2 T_1$.
This suggests that pure dephasing rates are relatively low, which is expected from the low photon occupation of the higher-frequency readout resonators.
The $T_2$ values, which add an additional $\pi$ pulse to decouple the qubit from low frequency noise, are not significantly different from $T_2^*$.
For qubits with lower $T_1$ values (such as qubit B4), the Hahn echo reduces the visibility of qubit oscillations to the extent that it is no longer possible to reliably measure $T_2$.
This implies that our qubit performance is primarily limited by dissipation ($T_1$) in the qubit and junction materials.

\section{Qubit Sensitivity to its Environment}
\begin{figure}[b!]
\centering
\includegraphics[width=3.3in]{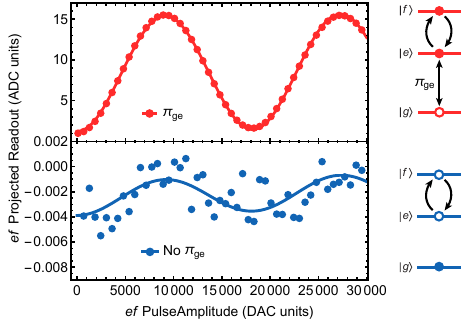}
\caption{
Qubit thermal population.
Residual $e$ state population of qubit C1 measured by comparing $ef$ Rabi oscillation amplitude with \textit{g} and \textit{e} swapped (pink) and idle populations (blue).
From this measurement, we find a probability of \qty{0.017\pm0.004}{\percent} of finding the qubit in its excited state.
\label{fig:fig3b}}
\end{figure}
A key benefit of higher-frequency qubits is a reduced thermal occupation of higher energy levels, reducing initialization errors and minimizing the need for active reset \cite{han2023active,Magnard2018reset}.
We measure the residual excited state population following the method in Ref. \cite{jin2015thermal}, as shown in \Fref{fig:fig3b}.
We perform Rabi oscillations between the \textit{e} and \textit{f} states by applying a Gaussian pulse of varying amplitude at the transition frequency $f_{ef}$ (see \Fref{fig:fig2}a). 
This measurement is also repeated with the $g$ and $e$ populations swapped with a $ge$ pi pulse immediately prior, resulting in a significantly stronger signal.
For maximum contrast, the readout signal is projected to resolve the \textit{e} and \textit{f} states (rather than \textit{ge}) for both measurements.
By comparing the relative oscillation amplitudes, we obtain a thermal excited state population of \qty{0.017\pm0.004}{\percent}, which is nearly an order of magnitude lower than values measured for microwave devices \cite{jin2015thermal}.
We note that the precision of this measurement is limited by our readout fidelity, which could be greatly improved by a K band quantum-limited amplifier \cite{Shu2021kpa,Banys2020kpa}.
This population corresponds to an effective qubit temperature \cite{jin2015thermal} of 116~mK, which is significantly higher than the physical fridge temperature, suggesting that improved filtering \cite{Corcoles2011filter,Wang2019cavAtten} could help reduce these qubit initialization error rates even further.

\subsection{Thermal Dependence of Coherence}
\begin{figure}
\centering
\includegraphics[width=3.05in]{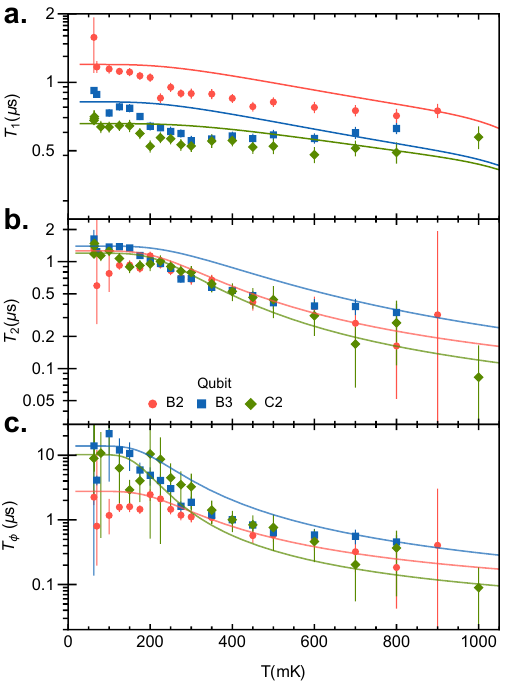}
\caption{
Thermal decoherence and dephasing.
(a) Decoherence time $T_1$ of three representative qubits measured as a function of temperature. 
A mild decrease is observed at higher temperature, consistent with a model including loss from increased system bath temperature (solid lines).
(b) Ramsey dephasing time $T_2^*$ as a function of temperature. 
The behavior is largely captured by a parameter-free thermal dephasing model assuming a fixed $T_1$ (solid lines).
(c) Pure dephasing rate $\Gamma_\phi$ which has dephasing from relaxation subtracted, resulting in better agreement with the model.
\label{fig:fig4}}
\end{figure}

To probe the thermal resilience of our qubits, we investigate our qubits at increased operating temperatures, shown in \Fref{fig:fig4}a. 
We observe a small decrease in $T_1$ with temperature above 300~mK, consistent with qubit heating from its environment \cite{lisenfeld2007nb2n}, but importantly we don’t see the drastic temperature dependence seen in qubits with aluminum junctions from quasiparticle-induced loss \cite{catelani2011quasiparticle,catelani2014quasiparticle,Serniak2018hotquasiparticle}, in line with expectations for niobium.
The measured dephasing time shown in \Fref{fig:fig4}b doesn't significantly decrease until above 200~mK, and is captured with the following parameter-free model for dephasing from thermal photons in the readout resonator \cite{Clerk2007dephasing,Reagor2016ms}:
\begin{equation}
T_\phi^{-1}=\Gamma_\phi=\frac{\gamma}{2}\text{Re}\left[\sqrt{\left(1 + \frac{2 i \chi}{\gamma}\right)^2 + \frac{8i\chi}{\gamma}n_\text{th}}-1\right]
\label{eq:tphi}
\end{equation}
Here $\chi$ is the dispersive coupling, $\gamma$ is the decay rate of the readout resonator, and $n_\text{th}=1/(e^{h f_R/kT}-1)$ is the resonator thermal population, set by resonator's fundamental frequency $f_r$.
The measured dephasing rate is then given by $T_2^{-1} = T_\phi^{-1} + T_{2,0}^{-1} + 1/(2 T_1)$, where $T_1$ and $T_{2,0}$ are measured by averaging the low-temperature values.
The model, which contains only independently-measured quantities, is overlaid for each qubit in \Fref{fig:fig4}b, showing relatively good agreement.
We can also go further by using the measured temperature-dependent values of $T_1$ to solve for $\Gamma_\phi$ directly, as shown in \Fref{fig:fig4}c, finding the the behavior is well modelled by \Eref{eq:tphi}.

\section{Conclusion}
By investigating transmon qubits at higher temperatures and frequencies than before, we gain a new perspective on superconducting qubit coherence.
Our devices highlight the importance of increased frequency, which could even help improve aluminum microwave device performance up to $\sim180$~mK, relaxing the strict thermalization requirements of current qubits \cite{Serniak2018hotquasiparticle,paik2011qubit3d,connolly2023nonequilib,Murch2012dephasing}, as well as exploring a new regime for studying physics with qubits in a conventional material platform.
Beyond this, we have shown that with high-temperature niobium trilayer Josephson junctions quasiparticle-based decoherence can be eliminated up to approximately 1~K.
Taking advantage of this, we confirm the thermal behavior of qubit dephasing unhindered by relaxation.
Above 200~mK, our qubit dephasing begins to see reduction below $1~\mu$s from readout-induced dephasing.
Already, these temperatures allow for significantly increased heat dissipation. 
This enables new opportunities for addressing thermal challenges in quantum processors, and new hybrid quantum experiments by reducing experimental complexity.
Moreover, expanding the achievable frequencies also expands the range of energy scales that can be modelled with superconducting qubits.
Niobium junctions can operate as high as 700~GHz: our devices have only begun to explore this accessible energy range, serving as an important first step in increasing qubit frequencies, and paving the way for even higher-frequency, higher-temperature superconducting quantum devices.

\section{Acknowledgements}
The authors thank P. Duda and K. H. Lee for assistance with fabrication development, F. Zhao for measurement support, and S. Anferov for useful discussions.
This work is supported by the U.S. Department of Energy Office of Science National Quantum Information Science Research Centers as part of the Q-NEXT center, and partially supported by the University of Chicago Materials Research Science and Engineering Center, which is funded by the National Science Foundation under Grant No. DMR-1420709.
This work made use of the Pritzker Nanofabrication Facility of the Institute for Molecular Engineering at the University of Chicago, which receives support from Soft and Hybrid Nanotechnology Experimental (SHyNE) Resource (NSF ECCS-2025633).



\appendix
\renewcommand{\thefigure}{S\arabic{figure}}

\setcounter{figure}{0}
\setcounter{section}{0}
\setcounter{equation}{0}
\renewcommand\theequation{\arabic{equation}}

\section{Fabrication Methods}
\label{appendix:fab}
\begin{figure*}[htb]
\centering
\includegraphics[width=6.67in]{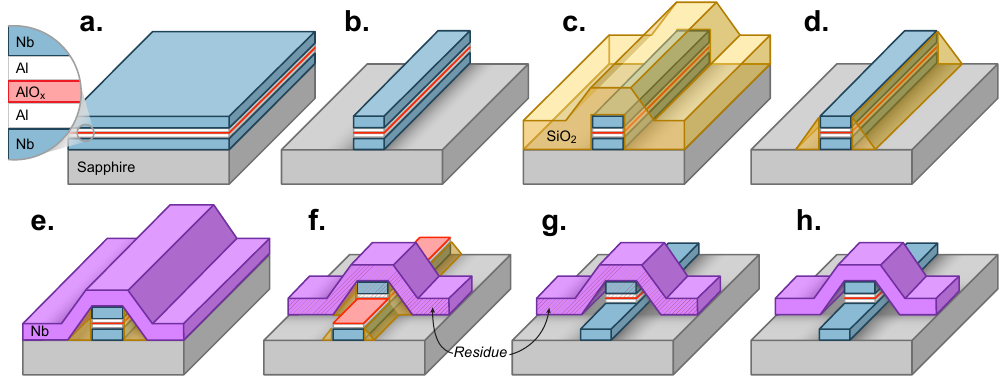}
\caption{
Junction fabrication process.
(a) Trilayer is deposited and oxidized in-situ. (b) First layer is etched with a chlorine RIE. (c) \ce{SiO2} is grown isotropically. (d) Sidewall spacer is formed by anisotropic etching with fluorine chemistry. (e) Surface oxides are cleaned in vacuum and wiring layer (purple) is deposited. (f) Second junction finger (and other circuit elements) are defined by a fluorine plasma etch selective against Al. (g) Final devices undergo a wet etch to further remove \ce{SiO2}, and exposed Al (h) 
Fluorcarbon residues are removed.
\label{fig:fab}}
\end{figure*}

\begin{table*}[htb]
\noindent
\begin{center}
\begin{tabular}{|c||c|c|c|c|c|c|c|c|c|c|c|c|}
\hline
&T(\qty{}{\degreeCelsius})&Pressure &
ICP/Bias Power& \ce{Cl2} & \ce{BCl3} &\ce{Ar} &\ce{CF4} &\ce{CHF3} &\ce{SF6} &\ce{O2} &etch time &etch rate\\
\hline
Etch 1 &$20\pm0.1$ &5 mT &\qty{400}{\watt} / \qty{50}{\watt} &30 &30 &10 &- &- &- &- &50-\qty{60}{\s}&$\sim\qty{4.5}{\nm/\s}$\\
Etch 2 &$20\pm0.1$ &30 mT &\qty{500}{\watt} / \qty{60}{\watt} &- &- &10 & 30& 20&- &- &120-\qty{140}{\s} s&$\sim\qty{2}{\nm/\s}$\\
Etch 3 &$20\pm0.1$ &5 mT &\qty{400}{\watt} / \qty{60}{\watt}  &- &- &10 &- &40 &20 &4 &65-\qty{90}{\s}&$\sim\qty{4.5}{\nm/\s}$\\
\hline
\end{tabular}
\end{center}
\caption{Plasma etch parameters used in the ICP-RIE etches described in the process. Etches are performed in an Apex SLR ICP etcher. Gas flows are listed in sccm.}
\label{tab:etches}
\end{table*}

\qty{330}{\um}-thick C-plane polished sapphire wafers grown with heat exchange method (HEMEX) are ultrasonically cleaned in toluene, acetone, methanol, isopropanol and de-ionized (DI) water, then etched in a piranha solution kept at \qty{40}{\degreeCelsius} for 2 minutes and rinsed with de-ionized water.
Immediately following, the wafers are loaded into a Plassys MEB550S electron-beam evaporation system, where they are baked at $>$\qty{200}{\degreeCelsius} under vacuum for an hour to help remove water and volatiles.
When a sufficiently low pressure is reached ($<\qty{5e-8}{\milli\Bar}$), titanium is electron-beam evaporated to bring the load lock pressure down even further.
The trilayer is now deposited by first evaporating \qty{80}{\nm} of Nb at $>\qty{0.5}{\nm}\text{/s}$ while rotating the substrate.
After cooling for a few minutes, \qty{8}{\nm} of aluminum is deposited while rotating the substrate at a shallow angle (10 degrees) to improve conformality.
The aluminum is lightly etched with a \qty{400}{\V}, \qty{15}{\mA} \ce{Ar+} beam for \qty{10}{\s}, then oxidized with a mixture of $15\%$ \ce{O2}:\ce{Ar} at a static pressure.
After pumping to below ($<\qty{e-7}{\milli\Bar}$), titanium is again used to bring the vacuum pressure down to the low $\qty{e-8}{\milli\Bar}$ range.
We note that the pressure for the remainder of the trilayer deposition is higher than for the first Nb layer.
The second \qty{3}{\nm} layer of Al is evaporated vertically while rotating the substrate to minimize void formation in the following layer.
The counterelectrode is then formed by evaporating \qty{150}{\nm} of Nb at $>\qty{0.5}{\nm}\text{/s}$.
The substrate is allowed to cool in vacuum for several minutes, and we attempt to form a thin protective coating of pure \ce{Nb2O5} by briefly oxidizing the top surface at $\qty{3}{\milli\Bar}$ for \qty{30}{\s}.

The wafers are mounted on a silicon handle wafer using AZ1518 photoresist cured at \qty{115}{\degreeCelsius}, then coated with \qty{1}{\um} of AZ MiR 703 photoresist and exposed with a \qty{375}{\nm} laser in a Heidelberg MLA150 direct-write system.
The assembly is hardened for etch resistance by a \qty{1}{\min} bake at \qty{115}{\degreeCelsius} then developed with AZ MIF 300, followed by a rinse in DI water.
The entire trilayer structure is now etched in a chlorine inductively coupled plasma reactive ion etcher (Etch 1 in Table \ref{tab:etches}).
The plasma conditions are optimized to be in the ballistic ion regime, which gives high etch rates with minimal re-deposition.
Immediately after exposure to air, the wafer is quenched in DI water: this helps prevent excess lateral aluminum etching by quickly diluting any surface \ce{HCl} (formed by adsorbed \ce{Cl} reacting with water vapor in the air).
The remaining photoresist is thoroughly dissolved in a mixture of \qty{80}{\degreeCelsius} n-methyl-2-pyrrolidone with a small addition of surfactants, which also removes the substrate from the handle wafer.

The wafer is ultrasonically cleaned with acetone and isopropanol, then \ce{SiO2} spacer is grown with low-temperature PECVD.
\ce{SiH4} and \ce{N2O} are reacted in a \qty{100}{\watt} plasma with the chamber held at \qty{190}{\degreeCelsius}.
The complete process (including chamber cleaning pumping and purging steps) takes approximately 15 minutes.
The wafer is then mounted on a silicon handle wafer using Crystalbond 509 adhesive softened at \qty{135}{\degreeCelsius}, then etched in a fluorine reactive ion etch (Etch 2 in Table \ref{tab:etches}).
This etch is optimized to be directional but in the diffusive regime to promote chemical selectivity while enabling the formation of the spacer structure.
At this point minimizing oxide formation is crucial since the top surface of the trilayer is exposed and will need to form a good contact to the wiring layer, so immediately following the completion of the etch, wafers are separated from the handle wafer by heating to \qty{135}{\degreeCelsius}, ultrasonically cleaned of remaining adhesive in \qty{40}{\degreeCelsius} acetone and isopropanol, then immediately placed under vacuum in the deposition chamber, where they are gently heated to \qty{50}{\degreeCelsius} for 30 min to remove remaining volatiles.

The contaminated and oxidized top surface of the counter electrode is etched with a \qty{400}{\V}, \qty{15}{\mA} \ce{Ar+} beam for \qty{5}{\min}, which is sufficient to remove any residual resistance from the contact.
After pumping to below ($<\qty{e-7}{\milli\Bar}$), titanium is used to bring the vacuum pressure down to the low $\qty{e-8}{\milli\Bar}$ range.
The wiring layer is now formed by evaporating \qty{160}{\nm} of Nb at $>\qty{0.5}{\nm}\text{/s}$.
The substrate is allowed to cool in vacuum for several minutes, and the wiring layer is briefly oxidized with $15\%$ \ce{O2}:\ce{Ar} at $\qty{3}{\milli\Bar}$ for \qty{30}{\s} to promote a thin protective coating of pure \ce{Nb2O5}.
The wafers are again mounted on a handle wafer, coated with AZ MiR 703 photoresist and exposed with a \qty{375}{\nm} laser.
The assembly is hardened for etch resistance by a \qty{1}{\min} bake at \qty{115}{\degreeCelsius} before development.
The final structure is now defined with a fluorine reactive ion etch (Etch 3 in Table \ref{tab:etches}).
This step proves to be highly problematic as it easily forms inert residues, and needs to be highly chemically selective in order to avoid etching through the aluminum, so the plasma is operated in a low-density ballistic regime with the addition of \ce{O2} which helps passivate exposed aluminum and increase selectivity.
To avoid the formation of excess sulfur silicon residues on the spacer (which we find are even worse than fluorocarbon residues), we reduce the \ce{SF6} content compared to Ref. \cite{anferov2023nbjj} and increase the Ar content to promote mechanical removal.
This instead favors the formation of fluorocarbon residue, which unlike the silicon residues can later successfully removed.
The etch time is calculated for each wafer based on visual confirmation when the bare wiring layer is etched through.
We remove crosslinked polymers from the photoresist surface with a mild \qty{180}{\watt} room temperature oxygen plasma that minimally oxidizes the exposed Nb (though find this is not very effective).
The remaining resist is now fully dissolved in \qty{80}{\degreeCelsius} n-methyl-2-pyrrolidone with surfactants.

\begin{figure*}[htp]
\centering
\includegraphics[width=5.2in]{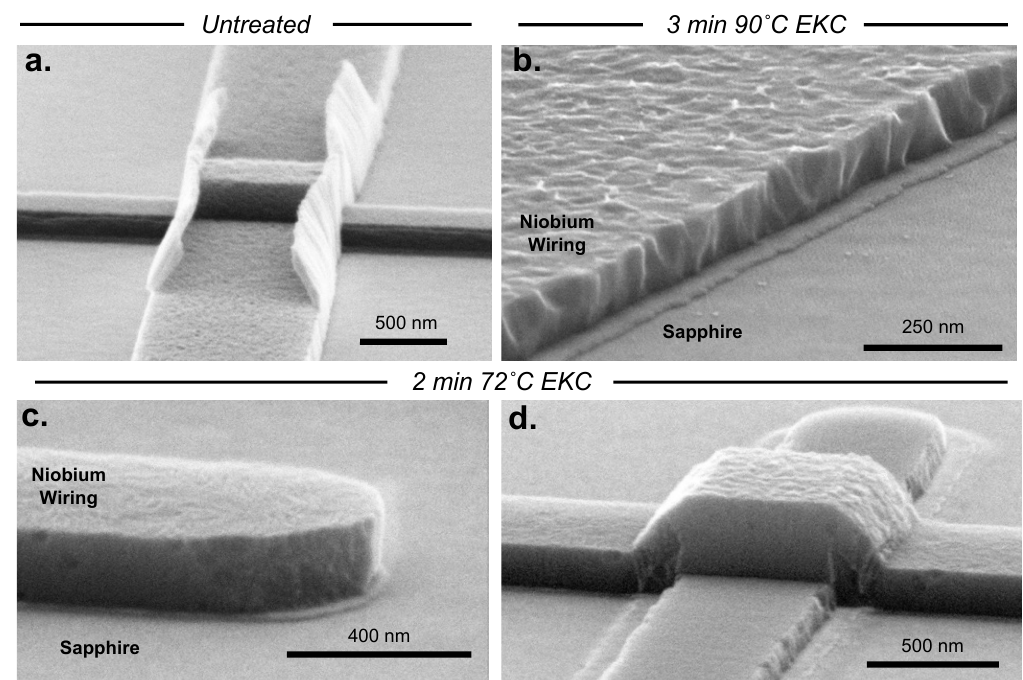}
\caption{
Fluorocarbon residue removal. 
(a) With the increased carbon content in Etch 3, the finished junctions have fairly pronounced fluorocarbon residues, with minimal spacer residues. 
(b) Finished junction treated with the EKC mixture nearly 20 degrees above the target etching temperature show significant metal attack (about 60~nm). Notably no sign of the fluorocarbon residues on the edges of the Nb wiring layer remain.
(c-d) When treated between 70--\qty{75}{\degreeCelsius}, the metal etch rate is reduced to a reasonable level, while the organometallic residue is still efficiently removed. This leaves incredibly smooth and virtually residue-free surfaces on the junction.
\label{fig:ekc}}
\end{figure*}
With the junctions now formed, the wafer is ultrasonically cleaned with acetone and isopropanol, coated with a thick protective covering of photoresist (MiR 703) cured at \qty{115}{\degreeCelsius}, and diced into \qty{7}{\mm} chips.
The protective covering is now dissolved in \qty{80}{\degreeCelsius} n-methyl-2-pyrrolidone with surfactants (we find this can also help remove stubborn organic residues from previous steps), and the chips are given a final ultrasonic clean with with acetone and isopropanol.
The remaining silicon spacer is now dissolved by a short 10-15$~$s etch in a mixture of ammonium fluoride and acetic acid (AlPAD Etch 639), quenched in de-ionized water, then carefully dried from isopropanol to preserve the now partially suspended wiring layer.
Finally we remove the aforementioned fluorocarbon residues by etching the chips in a \qty{72\pm1}{\degreeCelsius} solution containing hydroxylammine, catechol and 2-(2-aminoethylamine)-ethanol (Dupont EKC 265).
The finished chips are packaged and cooled down within 30 minutes from this final etch to minimize any \ce{NbO_x} regrowth from air exposure.

\section{Fluorocarbon Removal Methods}
\label{appendix:ekc}
As mentioned in Section \ref{appendix:fab}, the carbon and argon content of Etch 3 is increased to avoid lossy silicon spacer residues, which arise in sulfur-rich and carbon-poor plasma-etch chemistries \cite{thomas1987sulfur}.
Unfortunately this results in more pronounced fluocarbon residues \cite{anferov2023nbjj} on finished junctions, as shown in \Fref{fig:ekc}a.
As these residues are comprised of lossy dielectric material, it is essential to remove them to reduce junction loss.
To this end, we use commercially-available cleaning solution (Dupont EKC 265) specifically optimized to remove organometallic compounds.
This alkanolamine solution consists of a polar solvent (\ce{H2O}), hydroxylamine, 2-(2-aminoethylamine)-ethanol (AEEA), and catechol \cite{lee1996ekc,lee2004ekc}. 
The etch mechanism for this mixture begins with the reduction of any organometallic compounds by the hydroxylamine, allowing the compounds to become more soluble in the water and alkanolamine solution \cite{lee1996ekc,lee2004ekc}. 
This also reduces any exposed metal oxides, which conveniently also removes contaminated surface metal oxides, which are significant sources of loss \cite{verjauw2021nbOxide,premkumar2021nbOxide}.

The catechol, primarily a chelating agent or ligand, allows the now-soluble metal ions to form complexes and avoid precipitating out of solution. The hydroxylamine can also serve as a ligand in solution. Catechol also assists in metal protection and metal oxide solubilization due to its redox potential between 1.0~V to -2.0~V (relative to the standard hydrogen electrode) at a pH of 7 \cite{lee1996ekc}.
The alkanolamine of choice, AEEA, is an alcohol amine with a relatively high boiling point, high flash point, and nearly nonexistent metal or substrate etch rates under standard process conditions. The two-carbon linkage is key towards reducing attack on metals and alloyed substrates while enhancing reactivity with organic and organometallic residues.
Importantly, this mixture is only moderately acidic, with amines with hydroxyl groups both enhancing solubility and not increasing acidity greatly.
Reducing acidity may have the added benefit of reducing surface hydrogen content, which could help limit niobium hydride precipitation at cryogenic temperatures \cite{Ford2013hydride}, which is known to adversely affect niobium superconducting properties \cite{Jisrawi1998hydride} and coherence properties \cite{Ford2013hydride}.

\begin{figure*}[ht!]
\centering
\includegraphics[width=6.67in]{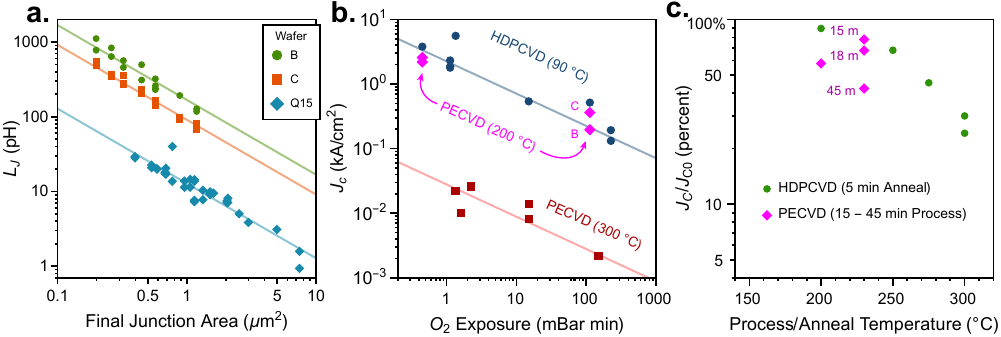}
\caption{
Low temperature PECVD junctions with high $J_c$.
(a) Junction inductance $L_J$ calculated from normal junction resistance $R_n$ measured as a function of junction area for several different devices with different process conditions, including junctions from wafers B and C (qubits measured in the main text).
A linear fit with respect to the junction area yields a measurement of the junction critical current density $J_c$
(b) $J_c$ of niobium trilayer junctions as a function of oxygen exposure (red and blue data from Ref. \cite{anferov2023nbjj}) with the addition of low temperature PECVD junctions (magenta), which still have high critical current density.
(c) Critical current density reduction as a function of anneal temperature, with the addition of the low-temperature PECVD junctions, which are only mildly annealed. 
The HDPCVD junctions (data from Ref. \cite{anferov2023nbjj}) were annealed for 5 minutes, while the junctions that went through PECVD spent more than 15 minutes at the temperature.
\label{fig:S1}}
\end{figure*}
As shown in \Fref{fig:ekc}b-d the mixture is effective at efficiently removing fluorocarbon residue.
At elevated temperatures (approaching the mixture boiling point) shown in \Fref{fig:ekc}b, the metal etch rate is increased, removing nearly 60~nm of niobium.
In practice this consumes too much of the junction, leaving behind very little of the first niobium layer, and leaves behind ridges, producing a rougher surface.
Reducing the etch temperature to 70--\qty{75}{\degreeCelsius} yields much more reasonable etch rates and surface profiles, as shown in \Fref{fig:ekc}c-d.
Only about 10~nm of niobium is consumed in the process, while the etch demonstrates selectivity towards the oranometallic residue materials (which can be up to 30~nm thick).
These etch conditions completely remove the fluorocarbon residues, and leave a very smooth and lightly etched finish on the junction metal surfaces.
Additionally, the mild surface etch ensures lossy surface interfaces and oxides \cite{wang2015psurf,verjauw2021nbOxide,premkumar2021nbOxide} are removed, which should improve loss characteristics in the device.

\section{High Critical Current Density with PECVD}
\label{appendix:pecvd}
Realizing qubits above 20~GHz requires Josephson junctions with critical currents higher than 0.1 kA/cm$^2$.
In the self-aligned niobium trilayer junction fabrication process \cite{anferov2023nbjj} this can be achieved by using low-temperature (\qty{90}{\degreeCelsius}) HDPCVD when depositing the spacer material, which avoids annealing the junction barrier \cite{morohashi1987nbjrev}.
Unlike PECVD, which requires hotter deposition temperatures, HDPCVD typically produces more porous material \cite{muraka2003dielectrics}: while the wet-etch rates are faster \cite{muraka2003dielectrics}, we empirically find that using HDPCVD spacers leads to worse junction loss properties (This may be a result of the rougher spacer surface producing a lossier metal-spacer interface left behind after the final wet etch).
Since the junction barrier annealing process is minimal below \qty{200}{\degreeCelsius}\cite{morohashi1987nbjrev,gronberg2017swaps,anferov2023nbjj} however, we can use PECVD at reduced temperatures (200--\qty{230}{\degreeCelsius}) while still forming junctions with a relatively high critical current density.
In this way by limiting the maximum temperatures and time spent at elevated temperatures, we can replicate the physical spacer properties of the junctions in Refs. \cite{anferov2023nbjj,gronberg2017swaps}.
This also improves process stability by providing control over a wide range of critical current densities with a unified process, eliminating the need for switching between PECVD and HDPCVD deposition methods.

\begin{figure*}[ht!]
\centering
\includegraphics[width=6.5in]{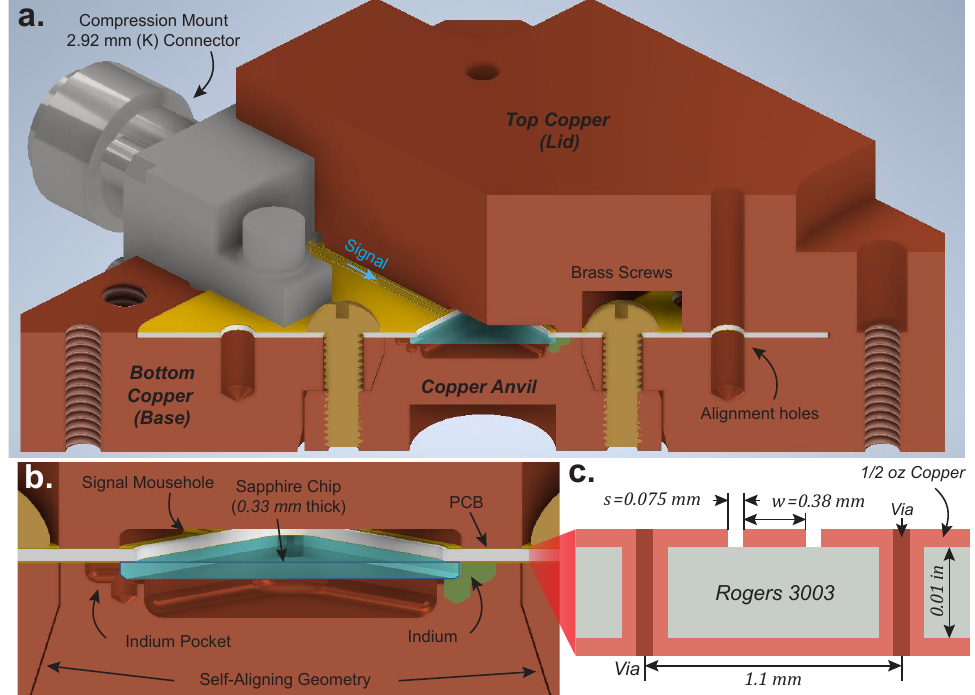}
\caption{
(a-b) Cutaway renderings of the K band packaging, highlighting the three machined copper pieces which secure, thermalize and electrically shield devices patterned on the sapphire chip (blue).
The sapphire chip is held in the anvil piece with sections of indium wire: as the anvil is screwed into the base, the anvil geometry aligns the chip within a tolerance of 0.005~inches and the indium wire (green) deforms into the machined pocket, firmly securing the sapphire.
(c) The package also routs a signal from K connectors to and from on-chip devices.
This is accomplished with a circuit board fabricated using a low-dielectric-constant substrate with good high-frequency loss properties.
Plated vias on either side of the signal trace combined with a mousehole slot in the copper lid ensure the signal is confined without excessive leakage and unwanted resonances.
\label{fig:S2}}
\end{figure*}
Using the relationship between the normal state resistance $R_n$ and the critical current $I_c$ \cite{ambegaokar1963relation}, we can use room temperature resistance measurements to predict cryogenic junction properties.
In \Fref{fig:S1}a, we show the Josephson inductance $L_J = \Phi_0/2\pi I_c$ determined from junction resistance as a function of final junction area (with lithographic reduction and etchback taken into account).
Solid lines show fits to an inverse function of area, which can be used to extract the effective critical current density $J_c$.
Along with test devices containing many junctions with large areas, we also plot witness junctions (with geometries identical to the qubits in the main text) fabricated alongside the qubits, which have different values of $J_c$ resulting from differing process conditions.
While the resistance of the qubit junctions is not directly measured, this gives a reasonable estimate of their junction inductance.
The junction critical current density can be easily adjusted by changing the oxygen exposure $E=P_{\text{O}_2}T$ (the product of oxidation time $T$ and oxygen partial pressure $P_{\text{O}_2}$) during the junction barrier formation step \cite{morohashi1987nbjrev}.
The critical current densities extracted by fitting junction resistance are shown as a function of oxygen exposure $E$ for varying process conditions in \Fref{fig:S1}b, overlaid with measurements from Ref. \cite{anferov2023nbjj}.
Our results (magenta) are consistent with the expected $J_c \propto E^{1/2}$ dependence, and importantly we find that reducing the PECVD deposition temperature from 300 to 200--\qty{230}{\degreeCelsius} results in critical current densities close to the low-temperature (\qty{90}{\degreeCelsius}) HDPCVD junctions.
Going further, we can directly compare the PECVD and HDCPVD processes to estimate the annealing effect of the elevated temperatures.
We summarize the reduction in $J_c$ as a function of the PECVD deposition temperature in \Fref{fig:S1}c, finding that our process typically maintains more than 50 percent of the HDPCVD critical current density.
These differences are comparable to the variations in $J_c$ measured from wafer to wafer (likely a result of imperfect process control), so we are not able to resolve a consistent trend with temperature \cite{anferov2023nbjj,morohashi1987nbjrev}.
However comparing the unintentionally different lengths of time each wafer spent at elevated temperatures (labelled in \Fref{fig:S1}c) could help explain the $J_c$ variations.
Automating and further refining the fabrication process described in Appendix \ref{appendix:fab} may help further increase junction consistency.
Nevertheless, within a practically accessible oxygen exposure, our junction process can realize critical current densities between 100~A/cm$^2$ and 2.5~kA/cm$^2$, while also taking advantage of the smoother and more conformal PECVD spacer surface.

\section{Low-loss K Band Packaging}
\label{appendix:packaging}
\begin{figure*}[htb]
\centering
\includegraphics[width=6in]{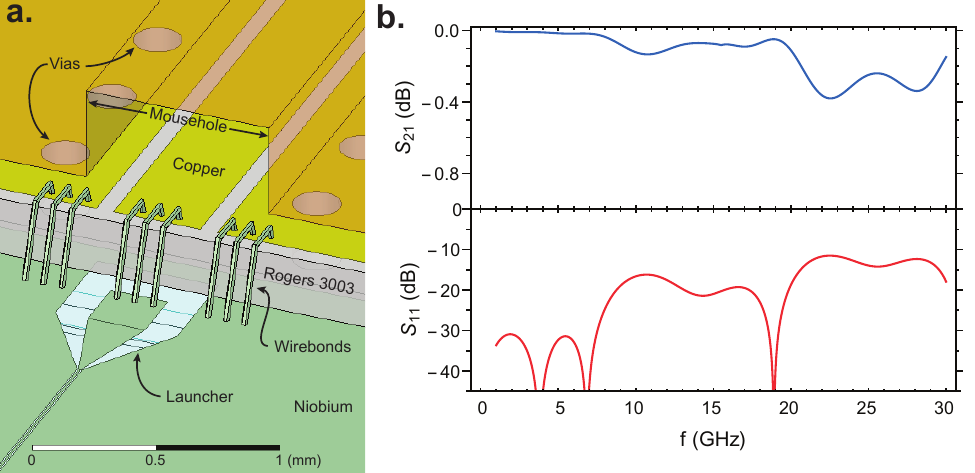}
\caption{
(a) Optimized launcher geometry and wirebond configuration which achieves maximal transmission up to 27~GHz. Using a manual wirebonder, we attempt to replicate this wirebond shape, however in practice the circuit board dimensions requires slightly longer bonds.
(b) Simulated transmission and reflection through the K band package containing a chip with a straight transmission line connected with the optimized launch and wirebond geometry.
Not taking into account metal dissipation and insertion loss from the circuit board connectors, the simulations predict a total insertion loss better than 0.2~dB up to 27~GHz for each transition.
\label{fig:wirebonds}}
\end{figure*}
Chips with several qubits and their readout resonators sharing a common microwave feedline are mounted in a K-band (18--27~GHz) package shown in \Fref{fig:S2}, which is carefully engineered for low-loss operation from DC to 30~GHz.
Inspired by microwave solutions \cite{huang2021packaging}, the package aims to shield the qubits from decoherence from external fields, thermalize the sapphire chip containing the qubits, while adding minimal unwanted resonant modes, and efficiently routing signals to and from the chip with minimal dissipation and reflection .
We accomplish this with a copper-clad porous ceramic-filled PTFE dielectric (Rogers 3003) circuit board patterned with via-fenced \cite{Bahl2003lumpedElements} coplanar waveguides for signal routing.
The sapphire substrate is clamped to the circuit board by a high-purity copper enclosure, as shown in \Fref{fig:S2}b.
This enclosure is composed of three precision machined pieces, which thermalize and mechanically secure the sample, and prevent leakage by fully enclosing the chip on all sides.
To minimize potential damage to the delicate surface of the chip from accidents during mounting, the chip is first placed into a slot on the copper anvil piece, where it sits on short sections of soft indium wire located on each corner.
The enclosure is designed to naturally align patterns on the chip surface with the circuit board: as the anvil is screwed into the copper base, the anvil geometry aligns the chip within a tolerance of 0.005~inches and the indium wire deforms into the machined pocket, firmly securing the sapphire.
When assembled, the sapphire substrate is suspended so that the surface qubits are well separated from the lossy copper surfaces \cite{huang2021packaging}, with the resulting dimensions optimized such that the nearest package mode lies above 27~GHz.
To further minimize potential loss contributions from the copper, the anvil and lid are polished and the copper oxide is etched with glutamic acid (flux) immediately prior to packaging. Further improvement could be achieved by plating the copper surface \cite{huang2021packaging}.

The signal is coupled on and off the sapphire substrate with wirebonds.
Since each of these has a high inherent inductance of about 1~nH/mm \cite{Wenner2011wirebond}, which is increasingly problematic at higher frequencies, we use sets of 3 wirebonds for each connection.
Using finite element method simulations (Ansys HFSS) we simulate and optimize the wirebond location and profile in conjunction with on-chip signal launch geometry for best performance in the K band.
The final geometry is illustrated in \Fref{fig:wirebonds}a. 
To estimate package performance and avoid unwanted resonant modes in lengths of transmission line, we also simulate the scattering parameters of the entire package, which includes the entire chip and circuit board.
The simulated parameters, shown in \Fref{fig:wirebonds}b suggest an insertion loss better than 0.2~dB up to 27~GHz, after which the impedance mismatch effects become increasingly pronounced.
While not visible in \Fref{fig:wirebonds}b simulations also suggest the presence of weakly-coupled modes localized in the substrate and indium mounting regions, typically between 25--29~GHz (depending on the exact geometry of the deformed indium).
Combined with reduced transmission at higher frequencies this determines an upper useful operating range of about 27~GHz for this packaging design.

\section{Cryogenic Measurement Setup}
\label{appendix:setup}
\begin{figure*}[h]
\centering
\includegraphics[width=5.5in]{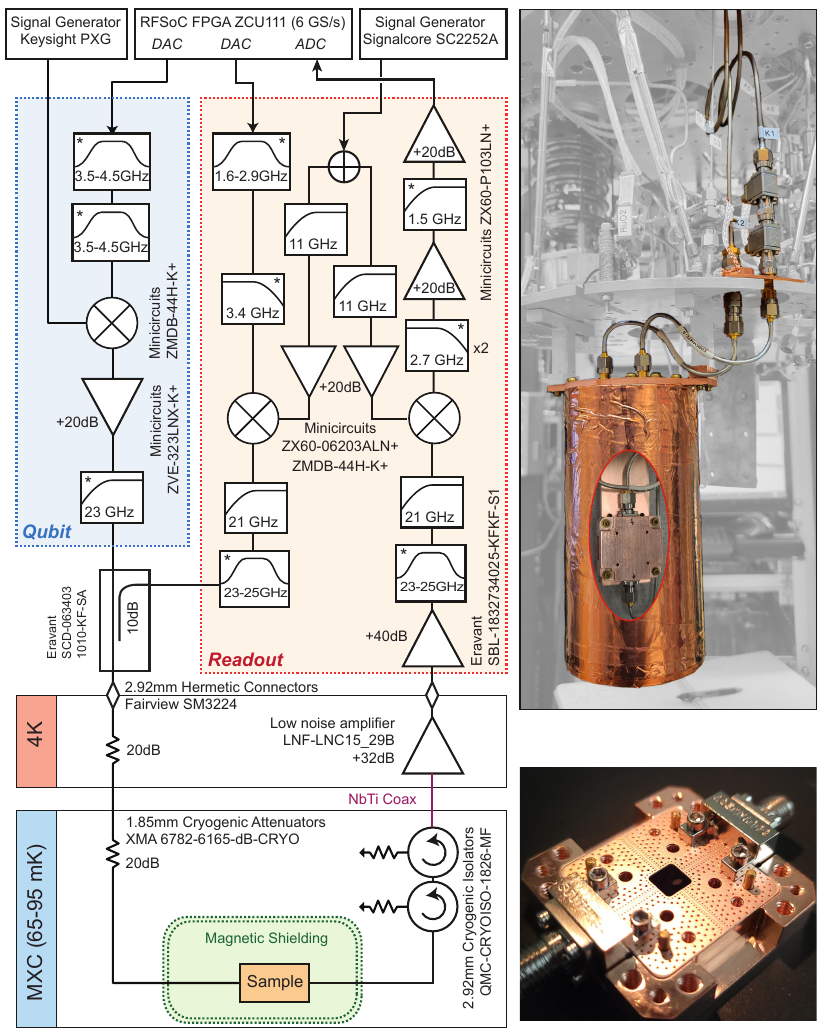}
\caption{
Schematic of the high-frequency microwave measurement setup used for qubit characterization. Colored tabs show temperature stages inside the dilution refrigerator. 
A composite photograph (top right) highlights the relevant hardware in the dilution refrigerator, with a cutaway showing the copper sample box location inside the magnetic shield.
A photograph (bottom right) shows the sample box with top and anvil copper pieces removed.
\label{fig:figS3}}
\end{figure*}
\begin{figure*}[ht!]
\centering
\includegraphics[width=6.5in]{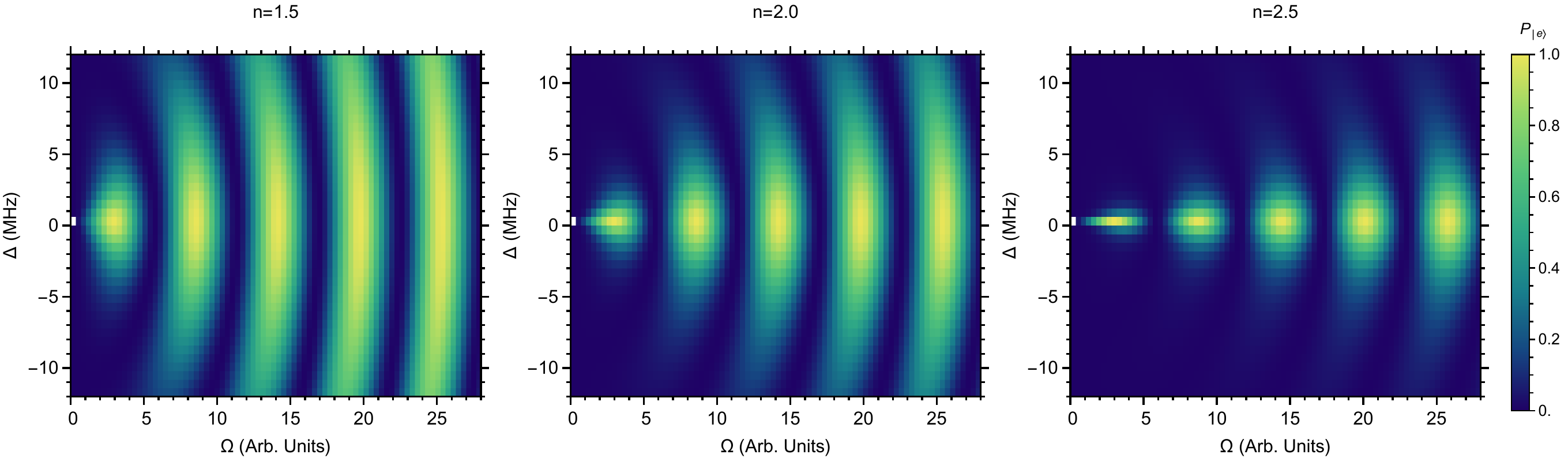}
\caption{
Excited state population following a finite-length Gaussian pulse terminated at $\pm n \sigma$, shown for different values of $n$.
For smaller $n$ the pulse shape is closer to a square pulse, and the fringes exhibit significant power broadening.
As more of the Gaussian envelope is used the fringe bandwidth is reduced, limiting the power broadening effect.
Here the pulse lengths $\sigma$ have been adjusted to yield similar oscillation rates at $\Delta=0$.
\label{fig:chevrons}}
\end{figure*}
The K band package described in Appendix \ref{appendix:packaging} is thermalized to the base stage of an Oxford Triton 200 dilution refrigerator (lowest mixing chamber temperature 65--95~mK), where transmission measurements through the central feedline can be used to characterize resonators and qubits.
The mounted assembly is encased in two layers of Mu-metal magnetic shielding to reduce decoherence from stray magnetic fields, as shown in \Fref{fig:figS3}.
The qubits are also protected from noise coming from the input line by 50 dB of cryogenic attenuation: these cryogenic attenuators maintain their properties up to excessively high frequencies (65~GHz) which helps ensure the noise temperature reaching the qubits is sufficiently low, even for higher frequency harmonics which will also be properly attenuated.
After passing through the sample, two cryogenic isolators along with a superconducting NbTi cable minimize signal dissipation while continuing to shield the qubits from thermal radiation.
A cryogenic low-noise amplifier (noise temperature around 8.5~K) combined with a low-noise room temperature amplifier help increase the output signal power to easily measurable levels while maintaining a good signal to noise ratio.

Resonators and qubit transitions are characterized with continuous-wave single and two-tone spectroscopy using a Agilent E5071C network analyzer (not shown in \Fref{fig:figS3}).
For pulsed qubit measurements, we use a Quantum Instrument Control Kit \cite{stefanazzi2022qick} based on the Xilinx RFSoC ZCU111 FPGA to synthesize and measure pulses, which are up and down-converted with the heterodyne measurement setup shown in \Fref{fig:figS3}.
The FPGA and carrier signal generators are clocked to a 10 MHz rubidium source for frequency stability.
Although the frequencies used can technically be sent through SMA connectors, in practice working at the edge of the connector operating band causes increased issues with impedance mismatches and unwanted circuit modes.
Since many components used in the setup are designed to operate at higher frequencies, we primarily use 2.92~mm (K) connectors for routing signals.
Since the individual filters (marked with stars in \Fref{fig:figS3}) have fairly wide bandwidths, we adjust the filter network to best match the frequency of each individual qubit in order to maximize transmission at qubit frequencies while filtering out leaking local oscillator signals and unwanted images.

\section{Rabi Oscillations with Finite Pulses}
\label{appendix:rabi}
An applied signal on-resonance with $f_{ge}$ will result in Rabi oscillations between the ground and excited state at a Rabi frequency $\Omega$ proportional to the signal amplitude.
In the main text, we explore this behavior by applying fixed-length Gaussian pulses near the qubit frequency which have the following pulse envelope function:
\begin{equation}
    \Omega(t) = \Omega_0 \exp\left[-\frac{t^2}{2\sigma^2}\right], \qquad -n \sigma < t < n \sigma
\end{equation}
For practical purposes, the pulse length is finite, which is achieved by truncating the Gaussian envelope at $\pm n\sigma$.
For simplicity, the resolution of the stored pulse waveform is limited to the FPGA processor speed, so we instead explore Rabi oscillations by varying the pulse amplitude (over which we have much finer control).
This behavior is sinusoidal, and results in periodic fringes with the qubit in its excited state, as shown in Fig. 2b in the main text.
For a continuous signal detuned from the qubit transition by $\Delta$, the excited state probability of the qubit would be given by \cite{Boradjiev2013finite}
\begin{equation}
    P_e (t) = \frac{\Omega^2}{\Omega^2 + \Delta^2}\sin^2(\pi \Omega t)
    \label{eq:pulseShape}
\end{equation}
The finite Gaussian nature of the pulse we use complicates the qubit evolution however, since the Rabi evolution rate is non-uniform during the pulse \cite{fischer2017finite} and the truncated pulse edges introduce non-adiabaticity in the qubit state evolution \cite{berman1998finite}.
For a truncated Gaussian pulse described in Eq. \ref{eq:pulseShape}, the excited state population can be modelled by \cite{Boradjiev2013finite,fischer2017finite}:
\begin{equation}
    P_e=\frac{\Omega^2e^{-n^2}}{\Omega^2e^{-n^2}+\Delta^2}\sin^2\left(\frac{\sigma}{2}\int_{-n}^{n}\partial\tau\sqrt{\Omega^2e^{-\tau^2}+\Delta^2 }\right)
\label{eq1}
\end{equation}

On resonance ($\Delta=0$) we recover the familiar expression $P_e=\sin^2\left(\pi\sigma\Omega\right)$. 
However off resonance the expected power-broadening of the Rabi oscillations in frequency space is reduced, as shown in \Fref{fig:chevrons}.
For small $n$ the behavior is similar to that of a square pulse \cite{Boradjiev2013finite}, however as we include more of the Gaussian profile by increasing the cutoff length $n$, the bandwidth of the oscillations decreases while the frequency scaling of the oscillations remains constant.
While this model neglects bandwidth effects on the pulse envelope reaching the qubit, the behavior shown in \Fref{fig:chevrons} qualitatively matches the behavior observed in Fig. 2b of the main text.

\section{Time Dendence of Coherence}
\begin{figure}[hbt]
\centering
\includegraphics[width=3.37in]{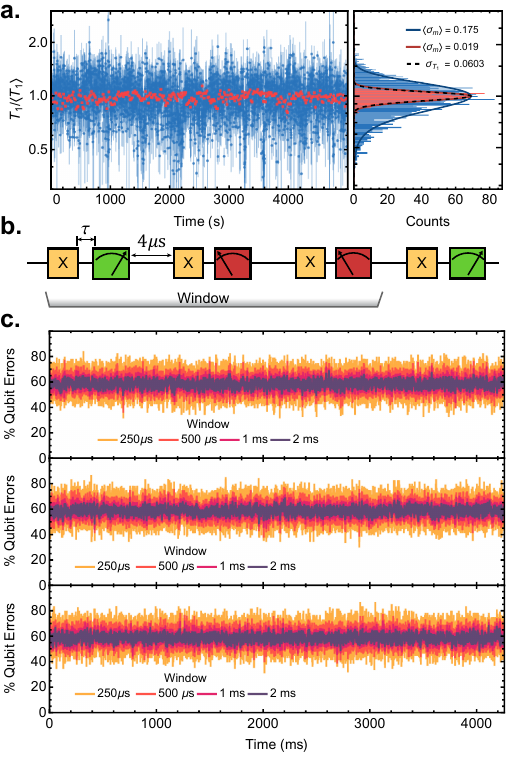}
\caption{
Temporal measurements with qubit C1.
(a) Decoherence time $T_1$ measured as a function of time, with low (blue) and higher average number experiments (red) interleaved.
Taking into account the individual fit errors, a cumulative histogram for both suggests the qubit $T_1$ has a variance of about 6\%.
(b) Rapid qubit error measurements are made by applying a pi pulse then measuring the qubit after a short delay $\tau$.
Experiments are repeated after a relaxation delay of $4~\mu$s, and the percentage of qubit errors are counted in a moving time window.
(c) Several time-dependent qubit error measurements with $10^6$ points, taken at different times, each shown with varying time window lengths.
Here the measurement is performed with a delay of $\tau=260~$ns, so a significant error baseline is observed from $T_1$ decay and limited readout fidelity.
We observe fluctuations consistent with small variations in $T_1$.
These measurements do not reveal conclusive signs of sustained qubit errors on a ms timescale such as those observed during cosmic ray events.
\label{figS4b}}
\end{figure}
\begin{figure*}[htb]
\centering
\includegraphics[width=6.2in]{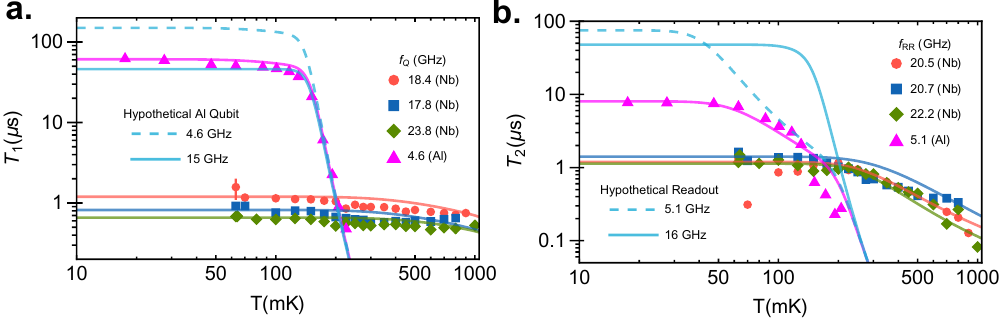}
\caption{
(a) Relaxation time for the K-band niobium qubits from the main text (labelled with qubit frequency) along with a microwave aluminum qubit \cite{anferov2023nbjj} (magenta), which showcases a much sharper drop in $T_1$ around 160~mK. Hypothetical aluminum-junction qubits with better performance are shown in cyan: we assume a fixed quality factor $Q_1$ which results in different base-temperature relaxation times, however the thermal dependence is unchanged.
(b) Dephasing time for the K-band and microwave qubits. In addition to the limitations of quasiparticle effects above 160~mK, the microwave qubit performance is degraded by excess thermal population in its readout resonator: this effect remains even if low-temperature performance is improved (dashed cyan). This dephasing can be mitigated by increasing the readout resonator frequency (solid cyan).
\label{fig:t1compareAl}}
\end{figure*}

\label{appendix:time}
Qubit coherence is expected to fluctuate in time, both from lossy coupled two-level-systems \cite{Carroll2022dynamics,Schlor2019fluctuators,Burnett2019benchmark} as well as non-equilibrium quasiparticles which can be generated by high-energy particle impacts \cite{Serniak2018hotquasiparticle,Serniak2019chargesensitive,mcewen2022impacts,Liu2024impacts,Wilen2021impacts}.
To investigate this, we measure coherence properties of a qubit as a function of time, as shown in \Fref{figS4b}a.
We interleave faster and slower experiments with $1000$ and $3\times10^5$ averages respectively, which each have different fit uncertainties $\sigma_m$.
After propagating out the respective average fit uncertainty $\langle\sigma_m\rangle$, we estimate that the qubit coherence fluctuation can be described by a Gaussian probability distribution with $\sigma_{T1}=\qty{6\pm1}{\percent}$ of the qubit $T_1$.
To speed up measurements, these population decay experiments are each performed with less than 10 delay points, however many averages are still needed to estimate the population decay rate.
Combined with data transfer times, the shortest experiment spacing achieved in \Fref{figS4b}a is around 3 seconds, which is sufficient to capture slow effects such as TLS population fluctuations \cite{Carroll2022dynamics,Schlor2019fluctuators,Burnett2019benchmark}.

Other dissipation processes such as quasiparticle relaxation \cite{connolly2023nonequilib,martinis2009nonequilib,serniak2019nonequilibrium} occur on a much faster timescale, especially in niobium \cite{leo2011nbquasiparticle}.
Since our devices are firmly in the transmon limit \cite{koch2007cpb}, we cannot leverage charge sensitivity to monitor quasiparticle dissipation dynamics through discrete charge fluctuations \cite{Serniak2019chargesensitive}.
To explore decoherence events with higher time resolution, we turn to rapid qubit error measurements similar to Refs. \cite{mcewen2022impacts,Liu2024impacts}.
Our measurement protocol (illustrated in \Fref{figS4b}b) consists of a deconstructed population decay experiment, in which the qubit is excited, and then measured following a short delay $\tau=260~$ns.
We wait for several coherence lengths ($4~\mu$s) in between experiments to ensure the qubit cools back down to its ground state.
The qubit state is measured (with readout fidelity 60~\%), and we count the percentage of errors occurring in a window of fixed time.
We note that due to hardware bandwidth limitations we are only able to measure a single qubit at a time, which limits the information gained by this measurement.

The error rates are summarized as a function of time using increasing window lengths in \Fref{figS4b}c.
Since we are measuring single-qubit errors and not correlated qubit errors \cite{mcewen2022impacts,Liu2024impacts}, our measurements cannot differentiate between different types of errors.
As a result the time-series data in \Fref{figS4b}c is dominated by a combination of readout errors and qubit errors from population decay.
An event generating an aluminum quasiparticle is expected to produce significant sustained increases in single and correlated qubit errors over durations of milliseconds \cite{Liu2024impacts,Wilen2021impacts,mcewen2022impacts}.
Other than small fluctuations (not inconsistent with $T_1$ distributions found in \Fref{figS4b}a), we do not observe conclusive signs of such millisecond-scale events. 
As niobium quasiparticles are expected to have lower equilibrium densities and much faster recombination rates on the order of 5~ns \cite{leo2011nbquasiparticle}, we cannot conclusively confirm or deny the presence or frequency of non-equilibrium quasiparticle events in our qubits. 
However when combined with the thermal dependence of coherence properties, we believe the absence of consistent qubit errors on millisecond timescales may suggest that our loss channels are primarily dominated by non-quasiparticle sources. 

\section{Qubit Thermal Performance in Context}
Our qubits benefit from both higher frequencies as well as a higher-energy gap superconductor, which increases the temperatures at which quasiparticle effects are observed.
As a comparison, it is useful to place their performance in context with a conventional microwave qubit: in \Fref{fig:t1compareAl} we show measured relaxation and dephasing times of a standard aluminum-junction microwave qubit (the same device compared to in Ref. \cite{anferov2023nbjj}) plotted alongside the devices discussed in the main text.

Qubit relaxation times are determined by qubit heating from its environment, which depends on the qubit frequency and environment temperature \cite{lisenfeld2007nb2n}, a low-temperature loss limit $T_{1,0}$ determined by independent sources of dissipation, as well as quasiparticle tunneling loss $T_{qp}$ \cite{catelani2011quasiparticle,catelani2014quasiparticle} resulting in the following model for temperature dependence:
\begin{equation}
    \frac{1}{T_1(T)} = \frac{1+ \coth\left(\frac{\hbar \omega_q}{2 k T}\right)}{T_{1,0}} + \frac{1}{T_{qp}(T)}
\end{equation}
Here the quasiparticle tunneling loss can be approximated as a function of temperature, qubit frequency $\omega_q$ and the superconducting energy gap of the junction $\Delta$ using Bessel functions $K_0$, $K_1$ as follows \cite{catelani2011quasiparticle,catelani2014quasiparticle,glazman2021quasiparticle}:
\begin{multline}
    T_{qp}\simeq\frac{4 \omega_q}{\pi}
    e^{-\Delta/k T}\cosh\left[\frac{\hbar\omega_q}{2kT}\right]\times\\\times\left(K_0\left[\frac{\hbar\omega_q}{2 kT}\right] + \frac{\hbar \omega}{4 \Delta}K_1\left[\frac{\hbar\omega}{2 kT}\right]\right)
    \label{eq:tqp}
\end{multline}
We find that this model is fairly accurate for describing qubit relaxation times for both niobium and aluminum junction qubits, as shown in \Fref{fig:t1compareAl}. As expected, the relaxation times of the aluminum junction qubit drop rapidly around 160~mK as the quasiparticle decoherence term becomes dominant.
From \Eref{eq:tqp} we observe that the niobium junction qubit would begin to see impact from quasiparticle tunneling contributions only above approximately 1.1~K.

Unlike the relaxation time, qubit dephasing times largely depend on the readout resonator properties. As described in the main text, the dephasing time has contributions from relaxation, along with pure dephasing from both the readout resonator and environment:
\begin{equation}    
\frac{1}{T_2} = \frac{1}{T_\phi} + \frac{1}{T_{2,0}} + \frac{1}{2 T_1}
\end{equation}
\begin{figure*}[htb]
\centering
\includegraphics[width=6.6in]{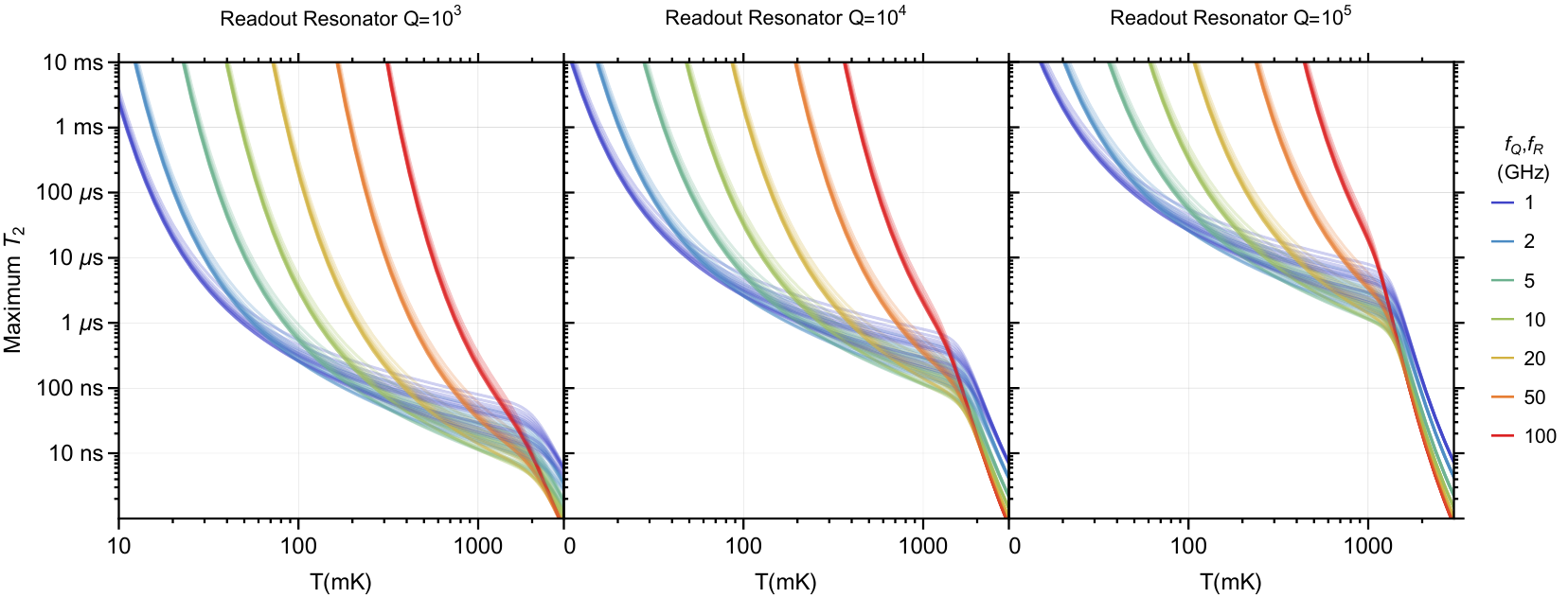}
\caption{
Maximum dephasing for all-niobium qubits, considering only thermal dephasing from the readout resonator and relaxation from quasiparticle tunneling. 
For a better comparison between frequencies, the readout resonator linewidth $\gamma$ is determined by the unitless quality factor $Q=\omega_R/\gamma$, and dispersive shifts $\chi$ between $0.3\gamma-5\gamma$ are considered. 
From this model, an increase in qubit and readout resonator frequency will result in improved dephasing times, particularly at lower temperatures.
Higher frequencies also increase the maximum operating temperature, with frequencies on the order of 100~GHz increasing dephasing performance near temperatures of 1~K.
\label{fig:maxdephasing}}
\end{figure*}
For the devices in the main text, the relaxation time is fairly constant so we primarily observe the effects of pure dephasing.
However for the aluminum qubit, shown in \Fref{fig:t1compareAl}b, we observe three regimes in which the qubit is first dominated by fairly constant dephasing from the environment, then reduced by contributions from the readout resonator above $\sim50~$mK, and finally above $\sim150~$mK is rapidly reduced by the drop in relaxation time caused by quasiparticle tunneling.
At the highest temperatures measured, the microwave qubit is sensitive to the linewidth of the readout resonator, which is in turn also likely increasing due to quasiparticle loss \cite{mattis1958bardeen} (which explains the deviations as we do not include this effect in the model).
As a result of the significant frequency difference, the aluminum qubit dephasing is comparable to the K-band qubits by approximately 140~mK.

This model of the thermal dependence of relaxation and dephasing leads to some interesting conclusions for qubit design.
Consider a hypothetical microwave qubit with improved performance ($T_{1,0}=T_{2,0}=150~\mu$s) and otherwise identical frequencies and properties as the measured microwave device (magenta).
Plotting this performance with dashed cyan lines in \Fref{fig:t1compareAl}a-b, we observe that the relaxation time behaves similarly to the measured device, with a sharp drop from quasiparticle tunneling around $150~$mK, while the dephasing times begin to decrease at temperatures as low as 50~mK.
Now consider the same hypothetical aluminum qubit, but scaled up approximately threefold in frequency, shown with a solid line.
Assuming the relaxation loss has material origins, the qubit quality factor will remain constant: this leads to a inverse reduction in relaxation time, with the thermal dependence largely unchanged.
Examining the thermal dependence of dephasing however, we conclude that an aluminum qubit merely three times higher in frequency should yield significant performance improvements between 100-200~mK!
Operating beyond these temperatures requires also addressing the quasiparticle loss through the use of a higher gap energy junction, as done with the qubits in the main text. 

Evidently, increased frequency can be a powerful tool for improving qubit performance at higher temperatures.
For the 20~GHz qubits described in the main text, we observe limits to the dephasing performance above approximately 250~mK.
Similar to the hypothetical example discussed above however, a further three-fold increase in niobium qubit frequency to 60--75~GHz could eliminate this thermal sensitivity and enable good performance at temperatures up to 1~K.
To illustrate this, we consider the maximum possible dephasing times for a niobium-junction qubit in \Fref{fig:maxdephasing}, where we assume a perfect device with $T_{1,0}=T_{2,0}=\infty$.
For any qubit design, quasiparticle-induced decoherence becomes dominant above approxiamtely 1.1~K.
The low-temperature dephasing performance is sensitive to qubit parameters however, and can be improved by decoupling the qubit from its readout resonator: achieved by either increasing the readout resonator quality factor $Q=\omega_R/\gamma$, or to a lesser extent by reducing the dispersive shift $\chi$.
However since this also reduces state measurement fidelity this approach is less practical.

On the other hand, increasing readout resonator frequency for a given qubit design leads to dramatic improvements, particularly for lower temperatures.
From \Fref{fig:maxdephasing} we also observe that as frequencies increase, so too does the maximum operating temperature (above which dephasing times decrease below a particular threshold).
From this we conclude that increasing the frequency of our K-band qubits to millimeter-wave frequencies (near 100~GHz) could further extend their operating temperatures from 250~mK to 1~K.

\bibliography{manuscript}
\end{document}